**Can a Tabula Recta provide security in the XXI century?**
By Francisco Ruiz, Associate Professor, Illinois Institute of Technology, Chicago. <ruiz@iit.edu>
2023

**Abstract**

In the not so unlikely scenario of total compromise of computers accessible to a group of users, they might be tempted to resort to human-computable "paper-and-pencil" cryptographic methods aided by a classic Tabula Recta, which helps to perform addition and subtraction directly with letters. But do these classic algorithms, or some new ones using the same simple tools, have any chance against computer-aided cryptanalysis? In this paper I discuss how some human-computable algorithms can indeed afford sufficient security in this situation, drawing conclusions from computer-based statistical analysis. Three kinds of algorithms are discussed: those that concentrate entropy from shared text sources, stream ciphers based on arithmetic of non-binary spaces, and hash-like algorithms that may be used to generate a password from a challenge text.

**Introduction**

Encryption is all around us. It has become one of the main pillars of our economy, as it ensures the confidentiality of data that is too sensitive to be exposed to prying eyes, such as bank transactions, medical records, and even video communications between individuals. Digital computers facilitate encryption by performing calculations that would take a long time to do with pencil and paper, to the point that encryption is no longer reserved to the leaders of armies and nations and every citizen can use it (though some national leaders, quite apparently, would prefer it if it would remain reserved to their exclusive use).

But encryption has used computers only for the last seventy years. Before then, we have many centuries where encryption—and decryption, its reverse—were indeed performed by individuals with hardly any equipment beyond a blank slate to write on. Paper-and-pencil cryptography, once the purview of kings, their spies, and counterspies, has been relegated to the status of a game for children, grown-up amateurs, and some rather deranged individuals.

Readers will excuse me if I prefer to use the term "human-computable" rather than "paper and pencil" to designate the same kind of cryptography. It seems to me that the latter term trivializes unnecessarily a field of research that has been of great consequence in the past and might still be useful in the future. Consider this not very unlikely scenario: your computer has been compromised by malicious hackers (which might work for a nation state actor, even your own) and you need to transmit a short piece of information to a friend, say, the location and code for a safe deposit box. Anything you type on the computer will be snatched and recorded immediately and will be sent to the enemy the moment to allow that computer to connect to Internet. If you have established a secret shared with your friend (say, a short passphrase), a human-computable cipher might save the day.

Likewise, I will not restrict "human-computable" to operations that a human can do strictly in the head without even simple aids like paper and pencil, as other researchers do (Rosenfeld, Vempala, and Blum, 2019). Such aids are ubiquitous, cheap, impossible to compromise via Internet, and easy to destroy after the job is done, yet they effectively supplement the very weak capacity of humans for remembering letters and numbers on a short-term basis. Computer-based algorithms are usually given as much memory as they want, and they are judged on the number of computational steps they require to complete the job. It is only fair to give humans a similar allowance, which in any case will be a lot smaller than what is given to even the weakest computer. In the case of this article, we will be using the traditional Tabula Recta, named originally by Trithemius, as a powerful aid for performing modular arithmetic operations directly with letters. Figure 1 shows a Tabula Recta (Latin for "Straight Table") for the 26-letter Latin alphabet used in English. Tabulae Rectae for other alphabets can be made easily by simply shifting the alphabet forward by one letter in each successive row. To add letters (modulo 26) using this aid, the user looks up the operands at the top and left headers and gets the result at the intersection of their respective row and column.

```
 |A B C D E F G H I J K L M N O P Q R S T U V W X Y Z
A|A B C D E F G H I J K L M N O P Q R S T U V W X Y Z
B|B C D E F G H I J K L M N O P Q R S T U V W X Y Z A
C|C D E F G H I J K L M N O P Q R S T U V W X Y Z A B
D|D E F G H I J K L M N O P Q R S T U V W X Y Z A B C
E|E F G H I J K L M N O P Q R S T U V W X Y Z A B C D
F|F G H I J K L M N O P Q R S T U V W X Y Z A B C D E
G|G H I J K L M N O P Q R S T U V W X Y Z A B C D E F
H|H I J K L M N O P Q R S T U V W X Y Z A B C D E F G
I|I J K L M N O P Q R S T U V W X Y Z A B C D E F G H
J|J K L M N O P Q R S T U V W X Y Z A B C D E F G H I
K|K L M N O P Q R S T U V W X Y Z A B C D E F G H I J
L|L M N O P Q R S T U V W X Y Z A B C D E F G H I J K
M|M N O P Q R S T U V W X Y Z A B C D E F G H I J K L
N|N O P Q R S T U V W X Y Z A B C D E F G H I J K L M
O|O P Q R S T U V W X Y Z A B C D E F G H I J K L M N
P|P Q R S T U V W X Y Z A B C D E F G H I J K L M N O
Q|Q R S T U V W X Y Z A B C D E F G H I J K L M N O P
R|R S T U V W X Y Z A B C D E F G H I J K L M N O P Q
S|S T U V W X Y Z A B C D E F G H I J K L M N O P Q R
T|T U V W X Y Z A B C D E F G H I J K L M N O P Q R S
U|U V W X Y Z A B C D E F G H I J K L M N O P Q R S T
V|V W X Y Z A B C D E F G H I J K L M N O P Q R S T U
W|W X Y Z A B C D E F G H I J K L M N O P Q R S T U V
X|X Y Z A B C D E F G H I J K L M N O P Q R S T U V W
Y|Y Z A B C D E F G H I J K L M N O P Q R S T U V W X
Z|Z A B C D E F G H I J K L M N O P Q R S T U V W X Y
```

*Figure 1 A classic Tabula Recta for the 26-letter Latin alphabet used in English*

The invention of the digital computer has changed ciphers forever, both from the defense (cryptography) and the attack (cryptanalysis) viewpoints. Human-computable ciphers now must contend with an attacker's ability to simply brute-force a solution by trying every possible key. Computer-enabled attackers can also perform sophisticated statistical analyses of the ciphertext that quickly reveal important metadata such as key length. If the cipher is based on functions that are invertible without excessive effort, a computerized attacker can find a solution in a very short time.

On the other hand, the same techniques can be used during the development of a human-computable cipher to make it more resistant to computer-based cryptanalysis. This is an area of research that has been largely abandoned in favor of strictly machine-computable ciphers, perhaps because this is where the money is. But human-computable ciphers have the possibility of beating current machine ciphers in certain areas. For instance, even the strongest of today's machine ciphers are typically a combination, repeated many times, of linear bit operations, which are intrinsically invertible, and nonlinear substitutions (s-boxes) based on a fixed pattern that has been carefully tested to remove weaknesses. The substitution subkey, if any, is added through a linear operation rather than a change in the substitution table.

**The Human-Computable Keyspace**

But before we get too far into developing human-computable cipher tools that may have any chance to stand against computer-aided cryptanalysis, we must determine whether its keyspace can ever be large enough to make computerized brute force difficult. Today (year 2023), a 128-bit symmetric key is still considered sufficiently long to prevent practical brute forcing, that is, a keyspace comprising $2^{128}$ = 3.4 x $10^{38}$ possibilities. Using alphabetic keys with 26 different characters, we would need log (3.4x$10^{38}$)/log (26) = 27.23 letters per key to get a keyspace of the same size. Twenty-seven letters is rather long for a single password (unless you write in German), but it is within reach for passphrases made of sentences.

Now, if the password or passphrase is intelligible because it is made of words contained in a dictionary, the real keyspace shrinks dramatically. Estimates of the actual information entropy of speech go as far back as Claude Shannon. He estimated the entropy of English text at 1.58 bits per character (Shannon, 1951). Therefore, a passphrase consisting of English words, 27 characters in length, contains an average of only 42.6 bits of entropy, a far cry from that of 27 random letters, and within the 40-bit realm of already-broken ciphers such as DES (Wobst, 2007). If text is to be used as a passphrase that a human can memorize, it is imperative, therefore, to take a large piece of text and then concentrate its entropy while reducing its length.

Fortunately, this is not particularly hard thanks to a property that all common speech has, and this is that its autocorrelation becomes weaker as the distance between sample points increases. Short-range autocorrelation is strong since there are many short bits of text such as "TH" or "QU" that are found very frequently compared to others, but as the distance between letters gets longer than the average length of a word (in English, sources quote anywhere from 5 to over 8 characters), the correlation fades gradually into what is expected for random letters. If we perform a simple operation with a pair of uncorrelated texts, the result is statistically closer to random than either of the original texts. For three uncorrelated English texts simply added mod 26, the result has a Shannon entropy of 4.67 bits on the average, quite close to the 4.7 bits of statistically random strings made of 26 different letters. And since English text shifted by more than 8 characters is essentially uncorrelated with itself, we just need 27 x 3 = 81 characters of common English text, split into three parts, then added together mod 26, to get the 27-letter string mentioned earlier that is quite close to random. These three parts have an average of 42.6 bits of entropy each. If the entropy is conserved when they are added together, we would get 127.8 bits of entropy, which is close to the target 128 bits.

**Ciphers using long pieces of text as key: TripleText and Snake**

The above consideration was made in reference to generating high-entropy keys for human-computable ciphers, but it can be used by itself as a cipher, which I call TripleText:
1. Sender Alice and recipient Bob agree on a shared corpus of conventional text, say, a book or a website with reliably consistent content (at least, between the time of the encipherment and that of the decipherment).
2. Alice decides a point to start drawing text (page + paragraph, or raw character count for a website), and takes a piece of contiguous text three times the length of the plaintext to be enciphered. Typically, spaces and punctuation are ignored.
3. Alice writes that text over three rows, plus the plaintext below them. Then she adds the resulting columns, modulo 26, using a Tabula Recta like that in Fig. 1.
4. The result is the ciphertext, which Alice transmits over insecure channels along with the position within the shared corpus where text began to be taken.

Recipient Bob gets the ciphertext and the location of the key source, and does the following:
1. Bob grabs the source and draws a piece of text three times the length of the ciphertext, starting at the location sent along with the ciphertext.
2. He then writes that text as three rows and adds them, modulo 26, using the Tabula Recta.
3. Then Bob subtracts the ciphertext from the sum of the three pieces of key text, using the same aid. To subtract letters using Tabula Recta, he looks up the subtrahend (the cipher, in this case) at the top header, goes down that column until the minuend (the key sum) is found, then left to read the resulting plaintext on the left header.

In essence, we are making a keystream by adding three pieces of hopefully uncorrelated text, and using that to encrypt the plaintext. It really does not matter whether the pieces of text are added or subtracted, so there is an opportunity for variations that can be carried out more easily on a Tabula Recta. A quicker form of TripleText, which I call "Snake" because the similarity of its movement over the Tabula Recta to a classic video game that goes by that name, replaces two-operand addition and subtraction with multiple-operand "serpentine" operations going this way: start with operand 1 at left, go right on that row until operand 2 is found, then up or down until operand 3 is found, and so forth, reading off the result at the corresponding header perpendicularly to the last movement. One can start equally well at the top header instead of the left, since the Tabula is symmetric around its main diagonal. An example is shown in Fig. 2.

*Figure 2 Serpentine operation H->C->A->R. Result: M*

Every time we change direction, we are subtracting the previous result from the last letter entered. Thus, the operation shown in Fig. 2 (H->C->A->R) is actually:

$-(-(-H + C) + A) + R \pmod{26} = -H + C - A + R \pmod{26} = M$

If we perform a serpentine operation with three pieces of key text K1, K2, K3, with the plaintext P placed at the top of the columns, the resulting ciphertext C will be, therefore:

$C = -(-(-P + K1) + K2) + K3 \pmod{26} = -P + K1 - K2 + K3 \pmod{26}$

So that, conversely:

$P = -C + K1 - K2 + K3 \pmod{26} = -(-(-C + K1) + K2) + K3 \pmod{26}$

Which is exactly the same process used for encryption, but starting from the ciphertext instead of the plaintext. It can be done most efficiently with serpentine operations starting from the ciphertext, then the first piece of key text, then the second, and the third, reading off the plaintext letter on the header perpendicular to the last motion. Readers are encouraged to run the example found in Appendix A and experience how quick the Snake cipher is. But is it secure?

Stream ciphers like TripleText and Snake range from trivially insecure like the Caesar cipher (all characters in the keystream are the same) to mathematically impossible to break like the Vernam cipher, which uses a truly random keystream. It all depends on how close to random the keystream is. Usable keystream "randomness" boils down to two criteria:

1. All single characters must appear with roughly the same frequency, given a large enough sample. Of course, if the sample length is not a multiple of 26, it is not possible for all English letters to appear with the exact same frequency, so a statistical test is needed. I typically use the Chi-squared test, which is more sensitive than Shannon entropy or Friedman's index of correlation. For strings composed of 26 different characters, the null hypothesis that all characters appear with the same frequency passes with 90% probability if the Chi-squared sum is less than 34.4. This Chi-squared sum is defined here as:

$$\chi^2 = \frac{\sum_{i=1}^{26}(f_i - L/26)^2}{L/26}$$

Where $L$ is the length of the sample and $f_i$ is the number of times a given letter appears in the sample.

2. Any two characters separated by a certain distance are no more likely than any other set of two characters separated by that same distance. This is equivalent to saying that characters in the keystream are independent. A Chi-squared test can be applied here as well, the null hypothesis being that they are independent. It passes with 90% probability if the following sum is less than 671:

$$\chi_d^2 = \frac{\sum_{i=1}^{26}\sum_{j=1}^{26}(f_{i,j} - f_i * f_j/26)^2}{f_i * f_j/26}$$

Where $f_i$, $f_j$ are the frequencies for given letters $i$ and $j$, respectively, and $f_{i,j}$ is number of times letter $j$ appears a certain number of characters $d$ after letter $i$.

Keystreams generated by combining three or more pieces of English text pass both tests easily. Not so with only two pieces, which is what we would get if we attempted to encrypt an English plaintext by adding to it another piece of English text, as in the classical Running Key cipher. Because the frequency distribution of English letters is not uniform (E is much more frequent than Z, and so forth) and because some two- and three-letter groups are more frequent than others (TH, QU, TION, etc.), cryptanalysts can find somewhere to pry and eventually separate the two combined texts (Bauer 2021). But the TripleText- and Snake-generated ciphertext is the combination of *four* pieces of text and passes stringent randomness tests, making it quite difficult to find a chink that can be exploited.

This is, of course, if users don't do silly things such as reusing keystreams. Security is maximized if the piece of the source text used to make the keystream is entirely different from message to message, rather than shifting the starting point by a few characters as one might be tempted to do, which would cause the keystream to repeat after a certain point for messages of the same length.

TripleText and Snake can be quite powerful in practice, because the true entropy of the keystream easily exceeds that of a conventional cipher, even if computed by machine, which cannot be larger than the entropy of its key—128 bits, 256 bits, or whatever it might be—even if the resulting keystream passes all tests of randomness. But users should be wary of the limited size of the source space. There are only so many book titles printed in the world every year, so many Web pages, and since the starting position for the keystream is revealed when it accompanies the ciphertext—unless it is sent within the previous encrypted message, with the danger of breaking communication if a message is lost—it is not inconceivable that a powerful computer might try millions of candidate publications until it finds the right one. If a rare source is used, such as a true one-time pad, it will stand out if an enemy undertakes a search of the sender's or recipient's living space. This has always been one of the liabilities that spies of the past had to assume when using one-time pads.

**Human-Computable Ciphers with Limited-Length Keys: FibonaRNG and PolyCrypt**

Can a more conventional cipher, requiring only a key of limited length, be designed for human computation, and still achieve security comparable to that of a machine cipher? In this case, the problem can be reduced to whether or not a human-computable pseudo-random number generator (PRNG) can be designed, having the desirable properties of statistical randomness, ease of computation, and non-invertibility. The operations that we will consider are whatever can be done on a Tabula Recta, that is, addition and subtraction modulo 26, and alphabetic substitution, using a one-to-one equivalence between the straight alphabet and another alphabet derived from a key, possibly supplemented by transpositions of different kinds.

Alphabetic substitution is very valuable for human-computable encryption, for it has these properties:
1. It is non-linear, like the s-boxes used in machine ciphers. This leads to difficulty in reversing an algorithm using substitutions along with other operations. One notable example is the "Zodiac 340" message, which took more than 50 years to decode even though its algorithm consisted of a mere substitution (special characters were used, but other than that no more difficult than a regular alphabet) combined with an unkeyed block transposition of fixed period (Goodin, 2020).
2. It has a large keyspace. For 26-letter alphabets, there are 26! = 4.03 x $10^{26}$ combinations, comparable to that of a 88.38-bit random binary key.
3. It can be implemented almost transparently into a Tabula Recta. The simplest way is to use alternative alphabets as headers to the table, and perform the substitution, or its inverse, while looking up inputs or obtaining outputs.

Transpositions take little time to do and are independent of other operations. They do not alter the non-uniformity of the character distribution of the plaintext, or whatever code they are applied to, however. Transpositions can be based on a key, which determines the order in which columns (initially written as rows) are to be read. Thus the keyspace for transpositions is of size N!, where N is the number of characters per row.

Additions and subtractions make it possible to combine two or more characters, with interesting results. The plaintext can be combined with itself, after a certain offset, resulting in Vigenère's classic Autokey cipher, or the same can be done on separate keystream material, or a combination of both keystream and plaintext. Adding ciphertext to the combination does not increase security since the ciphertext is available to attackers as well. Adding two keystream characters to obtain the next one constitutes a Lagged Fibonacci Generator (LFG) (Marsaglia, Zaman, Tsang, 1990). LFGs can be very good, having a period far longer than any conceivable plaintext, or bad with a period that repeats before the plaintext is ended, in which case it becomes a variant of the classic (so-called) Vigenère cipher, which is easy to break once the period length is determined. This has limited the use of LFGs in machine-computed PRNGs, which need to avoid bad cases because they are applied to very long plaintexts. But the requirement is not so strict for human computation since the plaintexts are going to be much shorter and bad cases can be spotted easily before they are used. On the plus side, LFGs based on addition or subtraction tend to produce statistical randomness of good quality, except for an unavoidable correlation between the two keystream characters used to generate a third, and that one, which may be manifested by not passing the second Chi-squared test described above when the distance between characters is the lag between parent and child elements in the keystream.

Therefore, here is another human-computable cipher, where the keystream is made with a LFG. I call it FibonaRNG:
1. Sender Alice and recipient Bob share three alphabetic secret keys: two are scrambled alphabets to be used for substitutions, the third is the seed of a LFG based on modulo 26 subtraction of letters, facilitated by a Tabula Recta. Subtraction is preferred over addition because it is easier to do on the Tabula Recta.
2. Alice places the first scrambled alphabet at the top of the Tabula Recta, and the second on the left side, where they become new headers.
3. Alice stretches the seed by as many letters as the plaintext has, to form the keystream, by performing this LFG algorithm: take two consecutive letters from the seed or keystream and look up the first at the top of the Tabula Recta, using the scrambled header so a substitution is performed automatically, then go down that column until the following letter is found, then to the left on that row, to read the keystream letter generated, thus performing another substitution automatically, with a different key from the first; mark the first letter used so it is not used anymore and go on.
4. Then Alice writes the plaintext above the keystream and does this for every plaintext letter and the keystream letter below it: look up the plaintext letter at the top, go down to find the keystream letter, then to the left to read the resulting ciphertext letter.

When Bob gets the ciphertext, he reproduces Alice's steps 1 through 3 in the same way, then writes the ciphertext above the keystream and does the equivalent of step 4, except that he begins looking up ciphertext letters on the left header, and reads off the plaintext letters on the top header.

As I was writing this article, I became aware of Frank Rubin's article in Cryptologia (Rubin, 1996), where he develops quite a similar algorithm starting from the premise of "Designing a High-Security Cipher" based on human-computable operations, which nevertheless he describes in bytes, as if it were meant to be performed by computer, rather than the original base 26 alphabet (or whatever alphabet the plaintext uses). Rubin calls the LFG operation "chained addition," and restricts it to a seed comprising 80 bytes. He uses the two substitutions I described, plus a third independent one applied to the keystream elements produced by the LFG. The FibonaRNG cipher I describe above uses substitution 2 both for this purpose and for the final combination of keystream and plaintext for greater simplicity, but I agree that these two substitutions do not have to be the same. A reader of this article might want to read the whole of Rubin's article as well since it presents different reasons to arrive at a very similar conclusion. Rubin focuses on resistance against computer-aided cryptanalysis, whereas I hope readers of this article will see I am trying to focus on usability without forgetting the need for strength.

A variant of the FibonaRNG cipher, which I ended up implementing in the open-source PassLok crypto suite, and therefore I named "PassLok Human Encryption", uses a randomly generated seed in step 3, which is combined in step 4 with a "seed mask" shared by Alice and Bob, and sent along with the ciphertext. Then Bob extracts the random seed before step 3, using the shared seed mask. This has the advantage that Alice and Bob can keep using the same set of shared keys, so long as a different random seed is used for each new message. An example is worked out in Appendix A.

A few comments about the security of this algorithm, which hopefully expands on Rubin's analysis of his quite similar cipher:
1. It is a stream cipher, but the combination of keystream and plaintext is not a mere addition or subtraction, but involves a substitution on the plaintext, and another substitution on the ciphertext. While the first substitution can be attacked by frequency analysis, the second cannot be attacked this way since the ciphertext passes randomness tests before the substitution. Ciphertext substitutions are hard to attack, in general, although some methods have been developed for short keys (Morelli and Walde, 2006).
2. The use of a substitution after the plaintext is combined with the keystream also combats the intrinsic malleability that stream machine ciphers tend to have unless accompanied by special check codes, because a linear change in the plaintext would lead to a non-linear change in the ciphertext.
3. If a random seed is used, the sender can keep reusing the same set of keys, because then every keystream is going to be different. The random seed needs to have good statistical properties (easy to do since its length will be the same as that of the seed mask, which likely will be short), which will remain after the seed mask is combined with it, even though the mask may be far from random.

4. On the other hand, a LFG-generated keystream is reversible in principle, and it is the unknown substitutions that make it hard to reverse. In a known-plaintext attack, only the substitutions would keep the attacker from learning the seed, and therefore the seed mask if the seed itself was random.
5. There's also the problem of errors during encryption. A human is much more likely to make a mistake performing an operation than a computer, even if aided by a Tabula Recta. While a mistake made when combining plaintext and keystream is confined to only one ciphertext character, an error in the LFG process corrupts the whole keystream, and therefore the whole ciphertext, after that point. But fortunately there is a method, based on partial cryptanalysis, that legitimate recipients can follow in order to overcome those mistakes. It is described in Appendix B.
6. A two-point independence test, as described above, applied to the keystream (whether with substitutions or without them) would come close to revealing the seed length, and a three-point test would absolutely reveal the dependence of every keystream letter on two consecutive letters some distance back, to give the seed length. Alternatively, an attacker that managed to obtain the keystream could look for repeated bigrams in that keystream—this happens, on the average, every 31 characters because of the birthday paradox—which would yield the same letter when the LFG operation is applied; this would reveal the seed length without any need for statistics. Just the existence of a deterministically repeated keystream letter would be a problem, for it would open the way to Kasiski analysis and all the other methods used against classic polyalphabetic ciphers.

The last problem mentioned can be reduced substantially by involving more than two (consecutive or not) letters in the LFG operation, replacing the simple subtraction with a serpentine operation on a Tabula Recta, with only a small increase in difficulty. These operations can involve two substitution keys: one on the first input, and one on output. A three-letter operation involving trigrams of consecutive keystream letters would be repeated every 148 characters, on the average. A four-letter operation with tetragrams would take more than 600 characters before it is seen again. So that quite likely no keystream tetragrams would ever be repeated within a message short enough to be encrypted by a human. I call this variant PolyCrypt, because it also has the ability to behave like a classic Vigenère cipher (only one keystream letter used to generate another) or even a simple substitution (no keystream letters used, leaving only the substitutions active). There is also an example in Appendix A.

But it might not be necessary for the sender to complicate the cipher this way. The keystream is only available in a known-plaintext scenario, and this only after the substitutions are undone. Even if the seed length is known, the distance between repeated keystream letters is not known (it would be different for different groups of specific letters, anyway), and this is what is needed for Kasiski analysis. When the statistical tests are applied to the ciphertext instead of the keystream, the correlation between ciphertext letters and pairs located a seed length before them fades to undetectable levels because the plaintext is in a way "encrypting" the keystream.

Rubin (1996) talks about a possible Depth Attack using multiple messages encrypted with the same keystream, which the substitutions still manage to thwart according to his analysis, but the scheme I present above does not lead to repeated keystreams if using random seeds instead of seeds directly derived form a key. Besides, he talks about a keystream of length over $10^{19}$ bytes, clearly meant for computer encipherment, whereas truly human-computed keystreams will reach 500 characters in length only rarely. In any case, as Rubin points out, the keystream generated by a LFG (or "chained addition", in his paper) depends on an internal state of length equal to the seed length, which can be long and, in the cipher I presented, also unknown to an attacker. This is potentially stronger than that of a machine cipher. We have seen earlier that 27 letters is equivalent to about 128 bits, 54 letters would be 256 bits, so long as the letters are random (the same, of course, goes for the bits), which they are, at least statistically, coming from a LFG. Therefore, a 54-letter seed (which does not even need to be random, or can be sufficiently masked by a 54-letter passphrase) would provide the same maximum keystream length as a 256-bit machine cipher, considered to be secure far into the future.

**Best Use of Transpositions and Plaintext Processing**

If Alice is truly paranoid, she will make an attacker's work even more difficult by adding a transposition. The best place for that is at the end of the encryption process because the ciphertext is already close to statistically random and anagramming it will give no clue. Were she to do it at the beginning, by transposing the plaintext, an attacker that got that far would have no difficulty extracting the plaintext since transpositions do not alter the statistical properties of the input, other than two-letter correlation. For best results, she should use a keyed transposition, where the input is written below the key by rows of the same length (except for the last possibly shorter row), the key is turned into numbers from 1 to key length using one of many possible schemes (alphabetical, for instance), and the input is read by columns starting from the one labeled 1, then 2, and so forth. This has the advantage that the total key space increases by a factor of N!, where N is the length of the transposition key, adding hardly any difficulty to the encryption process.

The one trick she could profitably do on the plaintext has been called by some "Russian copulation". Without entering on the etymology of the term, let me just say that it is pretty much the same as performing a "cut" on a deck of cards. Choose a random place to split the plaintext, and reverse the order of the two parts. If the spot for the cut is not evident from the context, place a marker word between them. Being so simple, this trick effectively combats known- and chosen- plaintext attacks, as well as message malleability, because the attacker would not know the location of the cut.

**Password generation using a Tabula Recta**

The third use of a Tabula Recta that I want to present is for generating passwords. All security experts on the planet recommend using passwords that are, (1) long, (2) different for different logins, (3) random-looking rather than made out of common words, if the user can handle it. They no longer recommend changing them frequently. All of this is because of the way a brute-force attack usually goes these days. Presumably there has been a previous compromise that has given attackers a database of hashed passwords, likely salted as well, so the only thing they need to do is hash whole dictionaries, which can be done quite fast even with commercially available GPUs, and look for matches in the stolen database. Once a match is found, the password has been found, and if the user is also identified all security is gone from other logins where that password has been recycled.

Password managers and passkeys increase security by storing login tokens (passkeys are more complex than simple passwords) that are different for different logins (recommendation 2), and can also generate long, random-looking passwords (recommendations 1 and 3), but they have problems of their own. They both involve supplementary software, which must be installed or be natively supported by the service we want to login into. They often—always, in the case of passkeys—involve a second electronic source of authentication, which can be stolen, lost, or corrupted. The elephant in the room that no one ever mentions is that they all introduce a third party that must be trusted.

Experts in cybersecurity have recognized these problems, and so have proposed alternative schemes to generate long, random-looking, and distinct login tokens without involving third parties, some of them based on human computation. Turing Award recipient Manuel Blum, to mention one, expounds the virtues of a human-computed challenge-response scheme (Cook, 2014) that roughly goes like this, with an example:
1. Come up with these two items that you must memorize: (1) a random mapping of the alphabet into the decimal digits 0-9 (a number of letters will map to the same digits, but that's okay), and (2) a random permutation of the digits 0-9. Together these two items constitute your secret key.
2. For each website you need to log in, decide what the "challenge" string will be. Most often it will be the name of the website. If that is too short, append some constant string to make it longer. It is okay if the string you append is always the same, so long as a hacker would have a tough time guessing it.
3. Now perform the following calculation in your head: Take the letters of the challenge and turn them all into digits using the secret alphabet mapping you memorized. For instance, the challenge for amazon.com is "amazon", which turns into these digits: 929073 because your alphabet mapping was "a" = 9, "m" = 2, and so forth. Now take the first digit, add it to the last, and keep only the last figure (9+3=12, so we keep 2), and then type in the digit occupying this place (10th for the 0) in your secret permutation (if the permutation was 5870163429, then the 2nd digit is 8). For the rest of the digits, add the last digit you typed to the next one in the challenge turned into numbers, and find the digit in the permutation occupying the position given by the last figure in the result (we obtain in succession 8+2=10 -> 9, 9+9=18 -> 4, 4+0=4 -> 0, 0+7=7 -> 3, 3+3=6 -> 6, for a final result: 894036).
4. If the website takes this directly, you're done. If it also requires lowercase, capitals, and so forth, append a ready-made string (fine if it's always the same) to appease the automatic password censor. So we may end up entering "894036aB$" to log into amazon.com.

The reason why this works is that a nonlinear operation (the alphabet mapping) is being applied for every input letter, and then another nonlinear operation (the digit permutation) for every output digit. If we call the letters $x_1$, $x_2$, and so forth all the way to $x_n$, the alphabet mapping $f(x)$, the permutation $g(x)$, and the output numbers $y_1$, $y_2$, etc., we have the following equations:

$$y_1 = g(f(x_1) + f(x_n) \bmod 10)$$
$$y_2 = g(f(x_2) + y_1 \bmod 10)$$
$$\ldots\ldots\ldots\ldots$$
$$y_n = g(f(x_n) + y_{n-1} \bmod 10)$$

This is quite similar to the LFG part of the FibonaRNG cipher described above, except that it is done in base 10 instead of base 26, and Blum recommends users to do the whole process in their heads. But there's no real need for users to convert the challenge to decimal digits and then do everything in their heads if they have a Tabula Recta in their pocket. They may even have a "Tabula Prava" where the headers have been replaced by randomized alphabets, comprising the secret key without any need for memorization.

This, then would be the password generation process using a Tabula Prava, where the top and left headers have been replaced by secret scrambled alphabets:
1. Take the first letter of the challenge (possibly the website's name) and look it up on the top alphabet. Go down that column until the last letter of the challenge is found, then go left to read the first letter of the password, which you type in.
2. Then take the next letter of the challenge, look it up at the top, and go down until you find the password letter you got on the previous step (it helps if you set the password entry box to visible), then go left to get the next password letter. Repeat this process, taking one new letter of the challenge each time until you get to the last one.
3. If the password is still too short, continue with a fixed sequence of extra letters as additional challenge text, which is bound to turn into something quite different for different logins. Finish typing in a fixed string made of capitals, digits, and special characters to appease the automatic censor.

This is essentially the same as Blum's method, except that users have no need to do any operations in their heads and nothing must be memorized. The secret codes remain written on the printed Tabula Prava, which is stored in a safe physical place. The keyspace for this method is actually larger than for the one with numbers: $26!^2$, equivalent to 177 bits of entropy. The resulting passwords, minus the stereotyped special characters appended, will consist of random-looking letters equivalent to 4.7 bits of entropy per letter.

Would an attacker who obtains a password so derived be able to compromise other passwords? This would involve coming up with the two mixed alphabets, which is like obtaining the functions $f(x)$ and $g(x)$ from the equations above, starting from a knowledge of the result and, presumably, the challenge text from which it is derived. This converts into a system of linear equations where the unknowns are $f(\text{``a''})$ to $f(\text{``z''})$ and $g^{-1}(\text{``a''})$ to $g^{-1}(\text{``z''})$, that is, 50 of them since the last letter in each alphabet is always forced by the others. So 50 equations are needed, which likely would require four or five compromised passwords, not just one. This would buy users the time needed to change the passwords in all their logins, starting from a different set of alphabets.

**Appendix A: Some Examples**

**Snake**

Let's say we want to encrypt the plaintext "ATTACK AT DAWN" (without spaces), and we have agreed to use Charles Dickens's "Oliver Twist" (gutenberg.org version) as the corpus shared with the recipients. We take the book and pick a random starting point for drawing text that will be written under the plaintext by rows of equal length as the plaintext. Because this is just an example, we choose the beginning of chapter 1 as starting position, and we take enough text to fill three rows below the plaintext, resulting in the following input:

```
ATTACKATDAWN
AMONGOTHERPU
BLICBUILDING
SINACERTAINT
```

Then we take the Tabula Recta and do this for each column of the above: look up the first letter at the top, go down to the second letter, then left or right to the third, up or down to the fourth, and then left to read the result on the left header. Example: A -> A -> B -> S, and then left to R. If a letter is repeated in the middle of an operation, we don't move and look for the letter that follows in the same direction we were going. Example: W -> P -> N -> N (no motion), and then left to T. The result, which we send along with a code telling recipients where we started drawing text from the book, is:

```
RQALFOCWYRTU
```

To decrypt this, the recipient places it at the top, and writes below the text from the book, resulting in this:

```
RQALFOCWYRTU
AMONGOTHERPU
BLICBUILDING
SINACERTAINT
```

Now, the same serpentine operations will give recipients the plaintext. Example: R -> A -> B -> S, and then left to A. Another: T -> P -> N -> N (no motion), and then left to W. Result:

```
ATTACKATDAWN
```

**PolyCrypt**

We will use three substitution keys and one mask for a random seed. The plaintext is still 'ATTACKATDAWN', and the keys are: WONDERFUL, MARVELOUS, and AWESOME; the seed mask is TERRIFIC. We begin by making substitution alphabets for the first three keys this way: start writing non-repeated letters in the key word; when a letter is repeated, write instead the one preceding it in the alphabet that has not been written yet (cycle from A to Z if necessary); finish by writing the remaining alphabet letters in reverse order. We obtain the following substitution alphabets:

```
1: WONDERFULZYXVTSQPMKJIHGCBA
2: MARVELOUSZYXWTQPNKJIHGFDCB
3: AWESOMDZYXVUTRQPNLKJIHGFCB
```

We write alphabet 1 at the top of the Tabula Recta, becoming the new top header, alphabet 2 at the left, becoming the new left header, alphabet 3 at right and bottom. The Tabula Recta becomes this "Tabula Prava":

```
      W O N D E R F U L Z Y X V T S Q P M K J I H G C B A
      -----------------------------------------------------
  M | A B C D E F G H I J K L M N O P Q R S T U V W X Y Z | A
  A | B C D E F G H I J K L M N O P Q R S T U V W X Y Z A | W
  R | C D E F G H I J K L M N O P Q R S T U V W X Y Z A B | E
  V | D E F G H I J K L M N O P Q R S T U V W X Y Z A B C | S
  E | E F G H I J K L M N O P Q R S T U V W X Y Z A B C D | O
  L | F G H I J K L M N O P Q R S T U V W X Y Z A B C D E | M
  O | G H I J K L M N O P Q R S T U V W X Y Z A B C D E F | D
  U | H I J K L M N O P Q R S T U V W X Y Z A B C D E F G | Z
  S | I J K L M N O P Q R S T U V W X Y Z A B C D E F G H | Y
  Z | J K L M N O P Q R S T U V W X Y Z A B C D E F G H I | X
  Y | K L M N O P Q R S T U V W X Y Z A B C D E F G H I J | V
  X | L M N O P Q R S T U V W X Y Z A B C D E F G H I J K | U
  W | M N O P Q R S T U V W X Y Z A B C D E F G H I J K L | T
  T | N O P Q R S T U V W X Y Z A B C D E F G H I J K L M | R
  Q | O P Q R S T U V W X Y Z A B C D E F G H I J K L M N | Q
  P | P Q R S T U V W X Y Z A B C D E F G H I J K L M N O | P
  N | Q R S T U V W X Y Z A B C D E F G H I J K L M N O P | N
  K | R S T U V W X Y Z A B C D E F G H I J K L M N O P Q | L
  J | S T U V W X Y Z A B C D E F G H I J K L M N O P Q R | K
  I | T U V W X Y Z A B C D E F G H I J K L M N O P Q R S | J
  H | U V W X Y Z A B C D E F G H I J K L M N O P Q R S T | I
  G | V W X Y Z A B C D E F G H I J K L M N O P Q R S T U | H
  F | W X Y Z A B C D E F G H I J K L M N O P Q R S T U V | G
  D | X Y Z A B C D E F G H I J K L M N O P Q R S T U V W | F
  C | Y Z A B C D E F G H I J K L M N O P Q R S T U V W X | C
  B | Z A B C D E F G H I J K L M N O P Q R S T U V W X Y | B
      -----------------------------------------------------
      A W E S O M D Z Y X V U T R Q P N L K J I H G F C B
```

*Figure 3. Tabula Recta with headers performing substitutions = "Tabula Prava"*

Now we prepare a new table consisting of three rows. On the second row, we write the seed mask, and then we go one position up and write the plaintext. Then we come up with some gibberish to fill the spaces above the seed mask, like this:

```
XGTSCVMU ATTACKATDAWN
TERRIFIC
(third row empty)
```

Now we fill the rest of the second row by taking a number of letters from the random seed and doing a serpentine operation starting at the top. We have agreed with the recipients to use three letters. Example: X -> G -> T, and read C at the bottom header. If a letter repeats, stay put and then move in the same direction we were going. When we run out of seed, continue with letters from the spaces below the plaintext. The result is this table:

```
XGTSCVMU ATTACKATDAWN
TERRIFIC CHFZQIBTHPFW
```
(third row empty)

The final operation consists of combining the columns, starting at the top, to get the ciphertext on the left, which we write on the third row. Example X -> T, and then read S at the left header. The result is:

```
XGTSCVMU ATTACKATDAWN
TERRIFIC CHFZQIBTHPFW
SSEVXIKG VHJMINROENLH
```

So the ciphertext we send is row three:

```
SSEVXIKGVHJMINROENLH
```

The recipients are going to do almost the same thing we did: (1) They write the ciphertext starting from position 1 on the top row, then the seed mask below it. (2) They extract the seed by combining first and second rows, starting with the *left side*, then to the mask letter, and reading the seed at the top; example: S (left) -> T -> X (top), and they write it on the bottom row. (3) They reconstruct the keystream from the seed following the same process as for encryption, and write it on the rest of the second row. (4) They get the plaintext by combining first and second row letters, but starting from the left side and ending at the top; example: V (left) -> C -> A (top). The resulting table looks like this:

```
SSEVXIKG VHJMINROENLH
TERRIFIC CHFZQIBTHPFW
XGTSCVMU ATTACKATDAWN
```

And the plaintext is everything following the random seed on the bottom row, that is:

```
ATTACKATDAWN
```

**Appendix B. Internet resources related to this article**

I have made Web pages describing the ciphers and tools discussed in this article (TripleText, Snake, FibonaRNG, PolyCrypt, PasswordPrava), having the advantage that the algorithms are also implemented in software (JavaScript), so readers can test them with a minimum of difficulty. Statistical tests are also built-in, and readers are encouraged to try their own plaintexts and keys, and see how close to random the keystreams and ciphertexts are.

The page for PolyCrypt includes a utility to overcome errors introduced during encryption, to be used by legitimate recipients who know all the keys. It is easy to spot where a mistake was made by looking at the decrypted plaintext, since the letter combined with the keystream letter where the error was made and most of those that follow will appear garbled. The solution is to introduce an intentional error at that spot during decryption. When the error introduced is the same that was made during encryption, a section of the decrypted plaintext will de-garble, up to the next error. The page has buttons to select the appropriate spot and save progress, so that eventually all encryption errors can be corrected.

Listed below are the links hosted on GitHub.com, which are expected to last a long time:

TripleText: https://fruiz500.github.io/ChaosFromOrder/TripleText.html

Snake: https://fruiz500.github.io/ChaosFromOrder/Snake.html

FibonaRNG: https://fruiz500.github.io/ChaosFromOrder/FibonaRNG.html

PolyCrypt: https://fruiz500.github.io/ChaosFromOrder/PolyCrypt.html

Password generation: https://fruiz500.github.io/ChaosFromOrder/PasswordPrava.html

More resources like these: https://github.com/fruiz500/chaosfromorder

Blog where these algorithms are explained in more detail, written for non-specialists: https://prgomez.com